\documentclass[]{spie}  
\usepackage[]{graphicx}
\pdfoutput=1

\def\arcsec{\hbox{$^{\prime\prime}$}}

\title{KAPAO: a MEMS-based\\natural guide star adaptive optics system} 

\author{Scott A. Severson\supit{a}, Philip I. Choi\supit{b}, Daniel S. Contreras\supit{b}, Blaine N. Gilbreth\supit{a}, Erik Littleton\supit{b}, Lorcan P. McGonigle\supit{b}, William A. Morrison\supit{b}, Alex R. Rudy\supit{b}, Jonathan R. Wong\supit{b}, Andrew  Xue\supit{c}, Erik Spjut\supit{c}, Christoph Baranec\supit{d}, and Reed Riddle\supit{d} 
\skiplinehalf
\supit{a}Sonoma State University, 1801 East Cotati Ave., Rohnert Park, CA 94928, USA; \\
\supit{b}Pomona College, 610 North College Ave., Claremont, CA 91711, USA; \\
\supit{c}Harvey Mudd College, 301 Platt Boulevard, Claremont, CA 91711, USA; \\
\supit{d}California Institute of Technology, 1200 E. California Blvd., Pasadena, CA, 91125, USA
}

\authorinfo{Further author information:\\S.A.S.: E-mail: scott.severson@sonoma.edu, Telephone: 1 707 664 2376\\  P.I.C.: E-mail: philip.choi@pomona.edu, Telephone: 1 909 607 0890}

\begin{document} 
  \maketitle 

\begin{abstract}
We describe KAPAO, our project to develop and deploy a low-cost, remote-access, natural guide star adaptive optics (AO) system for the Pomona College Table Mountain Observatory (TMO) 1-meter telescope. We use a commercially available 140-actuator BMC MEMS deformable mirror and a version of the Robo-AO control software developed by Caltech and IUCAA. We have structured our development around the rapid building and testing of a prototype system, KAPAO-Alpha, while simultaneously designing our more capable final system, KAPAO-Prime. The main differences between these systems are the prototype's reliance on off-the-shelf optics and a single visible-light science camera versus the final design's improved throughput and capabilities due to the use of custom optics and dual-band, visible and near-infrared imaging. In this paper, we present the instrument design and on-sky closed-loop testing of KAPAO-Alpha as well as our plans for KAPAO-Prime. The primarily undergraduate-education nature of our partner institutions, both public (Sonoma State University) and private (Pomona and Harvey Mudd Colleges), has enabled us to engage physics, astronomy, and engineering undergraduates in all phases of this project. This material is based upon work supported by the National Science Foundation under Grant No. 0960343.
\end{abstract}



\section{INTRODUCTION}
\label{sec:intro}  

The KAPAO project is an effort to produce a capable micro-electro-mechanical systems (MEMS) deformable mirror (DM) based astronomical adaptive optics (AO) system for the Pomona College Table Mountain Observatory (TMO) 1-meter telescope. Our project consists of a unique collaboration of private (Pomona College, PC, and Harvey Mudd College, HMC) and public (Sonoma State University, SSU) undergraduate institutions designing and building the system with the essential control software provided by our research university partner, California Institute of Technology (CIT), and their Robo-AO\cite{2012SPIE.8447E..04B} partner, the Inter-University Centre for Astronomy and Astrophysics. Our collaboration has produced a prototype system, KAPAO-Alpha, which has closed the AO loop on astronomical targets at the TMO 1-meter telescope. We present the design and the first-light results of our prototype, as well as an outline of our design for KAPAO-Prime, the facility class instrument that we will deploy later this year.

The inclusion of undergraduate student work in all aspects of this effort is integral to our process. The work presented herein includes \emph{significant} contributions from our undergraduate students within the framework of a carefully structured research environment. For a discussion of the design of this sort of undergraduate research see Severson 2010\cite{2010ASPC..436..449S}. Ultimately, the project goals of KAPAO can be enumerated as:

\begin{enumerate}
\item Science: Add high-resolution imaging capabilities to TMO, expanding current science programs with a focus on high-resolution, time-domain astrophysics.
\item Training: Introduce AO technology and techniques to a broad range of students; train a generation of undergraduates in both astronomical AO research and instrument design; fill future AO science and engineering pipelines.
\item Development platform: Use the completed facility-class AO instrument as an on-sky testbed for future undergraduates to explore expanded capabilities (e.g., coronagraphy, polarimetry and advanced wavefront sensing technologies).
\end{enumerate}

\begin{figure}[hb]
   \begin{center}
   \begin{tabular}{c}
   \includegraphics[height=8cm]{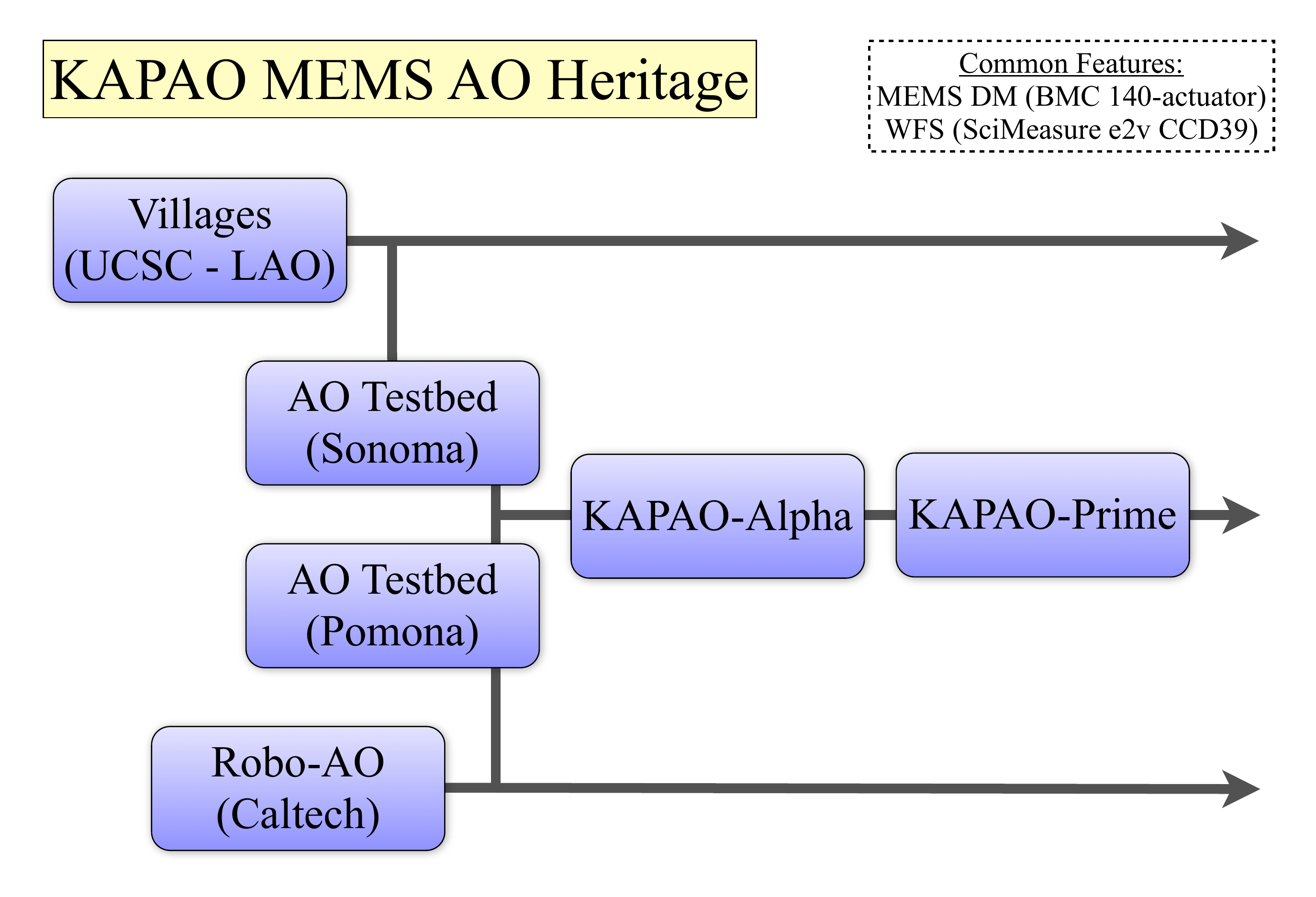}
   \end{tabular}
   \end{center}
   \caption[example] 
   { \label{fig:heritage} 
A presentation of KAPAO$'$s Heritage. 
Villages, a visible light AO system, built by the UCO/Lick Observatory Laboratory for Adaptive Optics, demonstrated the practicality of MEMS devices as astronomical AO deformable mirrors. Co-author Severson of the Villages group establishes AO testbed based on this architecture at Sonoma State University (SSU). At Pomona College (PC) co-author Choi establishes AO testbed for TMO 1-meter telescope. Co-authors Baranec and Riddle (Caltech)  build Robo AO, a robotic laser guide star MEMS DM AO system at Palomar. KAPAO-Alpha is a prototype system, designed and built at Pomona College with students from PC, SSU, and Harvey Mudd College (HMC) using control software from Caltech. We present first-light results with KAPAO-Alpha. KAPAO-Prime is the facility instrument for TMO and first light is expected in 2013.
}
   \end{figure}

Figure~\ref{fig:heritage} and caption describe the heritage of our KAPAO project. The Villages visible light MEMS DM based AO system
\cite{2008SPIE.6888E...4G}
\cite{2008SPIE.7015E...8G}
\cite{2008SPIE.7018E.126G}
\cite{2010SPIE.7736E..56M}
 was a key precursor to this work. The Robo-AO\cite{2012SPIE.8447E..04B} autonomous laser guide star AO system is a recently completed MEMS DM based system and provides the AO loop control software\cite{2012SPIE.8447E..2OR} to the KAPAO project. The consortium consisting of PC, SSU, HMC and CIT has completed the KAPAO-Alpha prototype and is building the KAPAO-Prime final instrument. All four of these systems are based on the same DM architecture, an 140-actuator BMC MEMS deformable mirror\cite{2010SPIE.7736E..80C}. For an in-depth discussion of issues relating to MEMS DMs for astronomical AO use, see Morzinski 2012\cite{2012SPIE.8253E...2M}.

\begin{figure}
   \begin{center}
   \begin{tabular}{c}
   \includegraphics[width=6.5in]{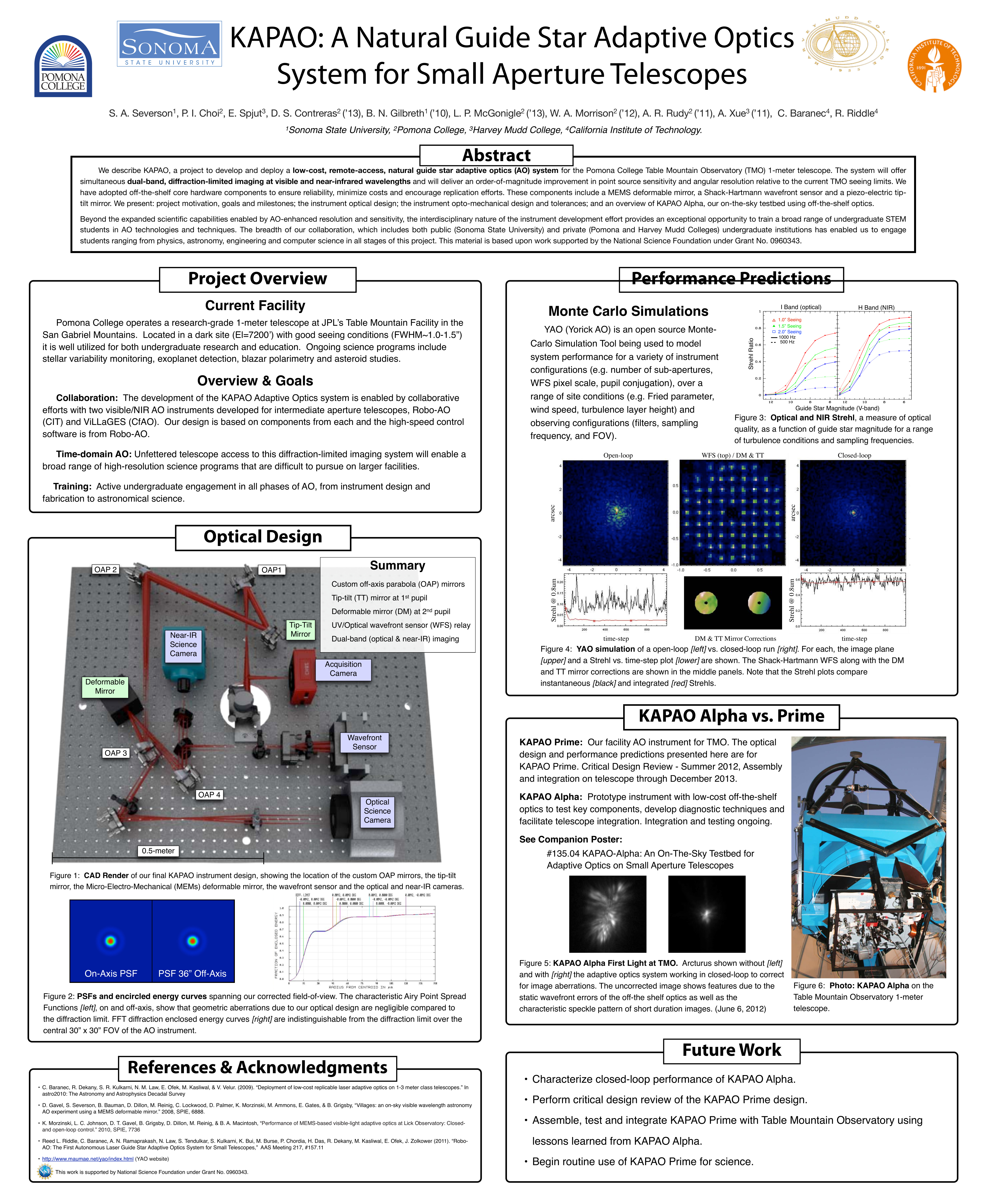}
   \end{tabular}
   \end{center}
   \caption[example] 
   { \label{fig:yao} 
A Yao Monte-Carlo simulation of an open-loop [left] vs.\ closed-loop run [right] with KAPAO at Table Mountain Observatory. For each simulation, the image plane [upper] and a Strehl vs.\ time-step plot [lower] are shown. Each time-step represents one measurement of the AO loop operating here at 1000 Hz. The simulated Shack-Hartmann WFS along with the DM and TT mirror corrections are shown in the middle panels. Note that the Strehl plots compare instantaneous [black] and integrated [red] Strehls.}
   \end{figure}

\begin{figure}[ht]
   \begin{center}
   \begin{tabular}{c}
   \includegraphics[width=6.5in]{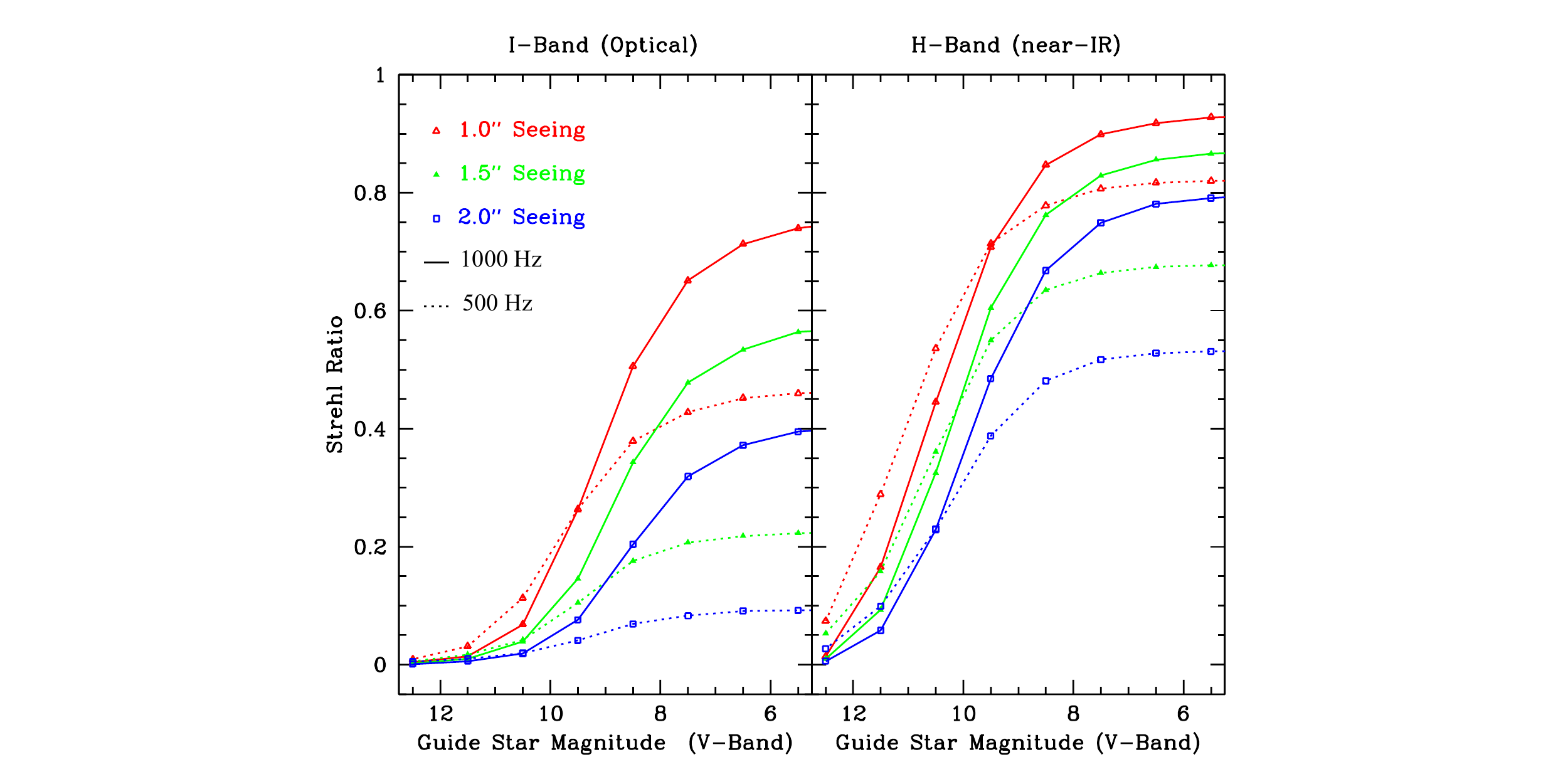}
   \end{tabular}
   \end{center}
   \caption[example] 
   { \label{fig:strehl} 
Simulated Strehl ratio vs.\ natural guide star V-band magnitude for KAPAO at the Table Mountain Telescope. The plot is divided into optical I-band strehl shown on the left and the near-infrared H-band strehl shown on the right. Both plots contain six curves; pairs of curves show loop operation at 1000 Hz [solid] and 500 Hz [dotted] WFS frame rates, under atmospheric seeing conditions of 1.0\arcsec [red large triangle], 1.5\arcsec [green small triangle], and 2.0\arcsec [blue square]. The system is demonstrably capable at visible and near-infrared wavelengths. For example, with typical seeing at TMO of about 1.5\arcsec, we calculate a limiting NGS magnitude of 10th magnitude for moderate IR correction.
}
   \end{figure}

With the feasibility of MEMS DM AO having been demonstrated, the KAPAO consortium produced simulations of the predicted performance of a system for the TMO 1-meter telescope. Figure~\ref{fig:yao} presents the results of the Yao Monte-Carlo simulation\cite{yao} of the performance of the complete system.
Figure~\ref{fig:strehl} presents the simulated Strehl ratio vs. natural guide star V-band magnitude for KAPAO. These served as the basis of our confidence in the KAPAO-Alpha and KAPAO-Prime designs and construction.

\graphicspath{{./images/}}

\graphicspath{{./images/}}

\graphicspath{{./images/}}

\section{KAPAO-Alpha DESIGN}
\label{sec:design}  

KAPAO-Alpha was conceived as a rapid-deployment prototype once we had early success in adapting the Robo-AO control software to our laboratory testbed system. Based on the same model BMC 12x12 actuator MEMS DM and a SciMeasure E2V-CCD39 wavefront sensor (WFS) as the Robo-AO system, we were able to close the AO loop with the testbed in summer 2011. With key questions regarding the mechanical coupling and software integration with the telescope as well as a preference to bring students through the process of designing and aligning optical systems designed around off-axis parabola (OAP) mirror relays, we split the development of KAPAO into the rapid design, construction and deployment of the KAPAO-Alpha prototype, while taking time to design KAPAO-Prime in careful detail.

\begin{figure}[]
   \begin{center}
   \begin{tabular}{c}
   \includegraphics[width=6.5in]{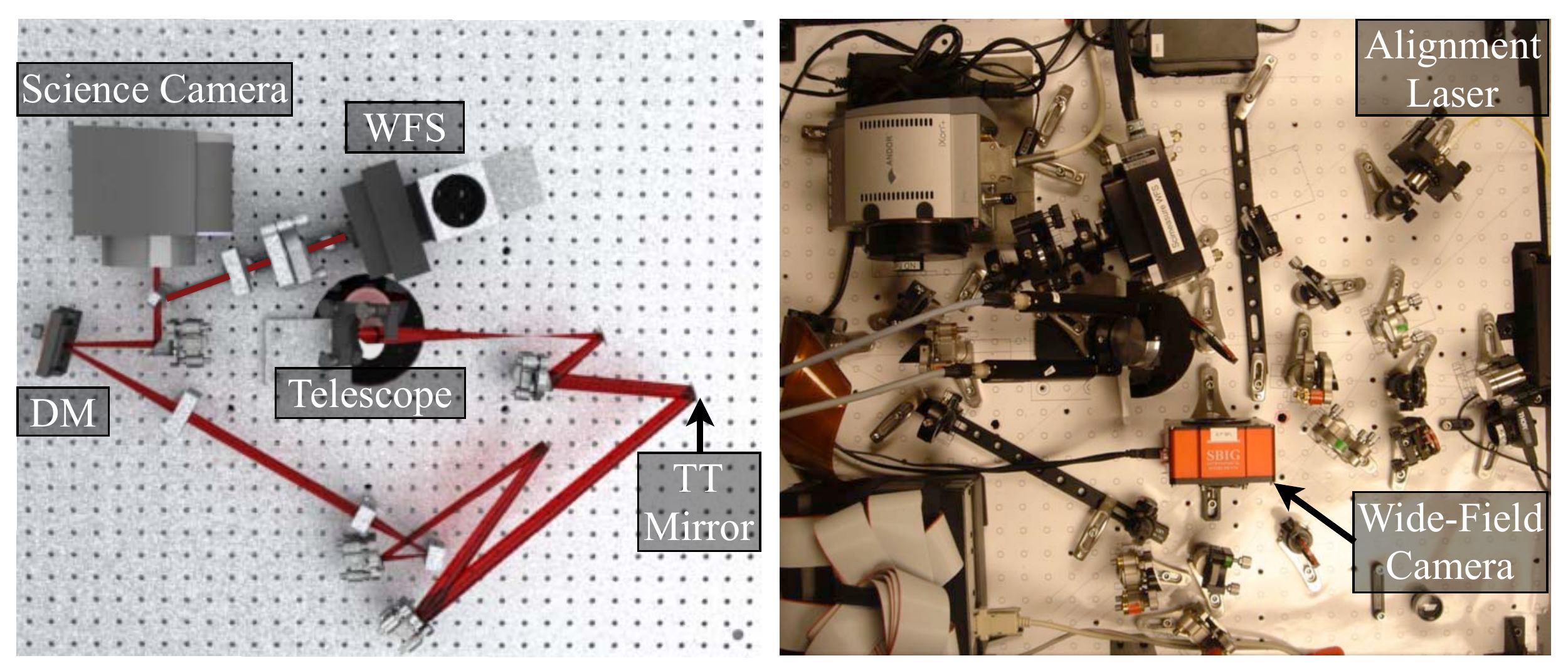}
   \end{tabular}
   \end{center}
   \caption[example] 
   { \label{fig:alpha} 
KAPAO-Alpha schematic layout [left] and as-built optical bench [right], shown in the same orientation. The optical bench is mounted to the telescope upside-down with the cassegrain focus folded [Telescope] into a plane parallel to the optical bench. The Tip-Tilt Mirror is downstream from the telescope focus, a fold mirror, and the first Off-Axis Parabola (OAP). The system used commercially-available OAPs and required a pupil expansion relay following the second OAP, a fold mirror, and a third OAP, before reaching the MEMS deformable mirror [DM]. After OAP4, a beamsplitter sends the beam to the Science Camera and the wavefront sensor [WFS]. When built [right], we added a beamsplitter and wide-field acquisition camera, as well as an alignment laser and an alignment camera. The system, while not optimized for throughput, was nonetheless a capable prototype for the final system, introducing our undergraduate students to a variety of optical and opto-mechanical methodologies. 
}
   \end{figure}

\begin{figure}
   \begin{center}
   \begin{tabular}{c}
   \includegraphics[width=6.5in]{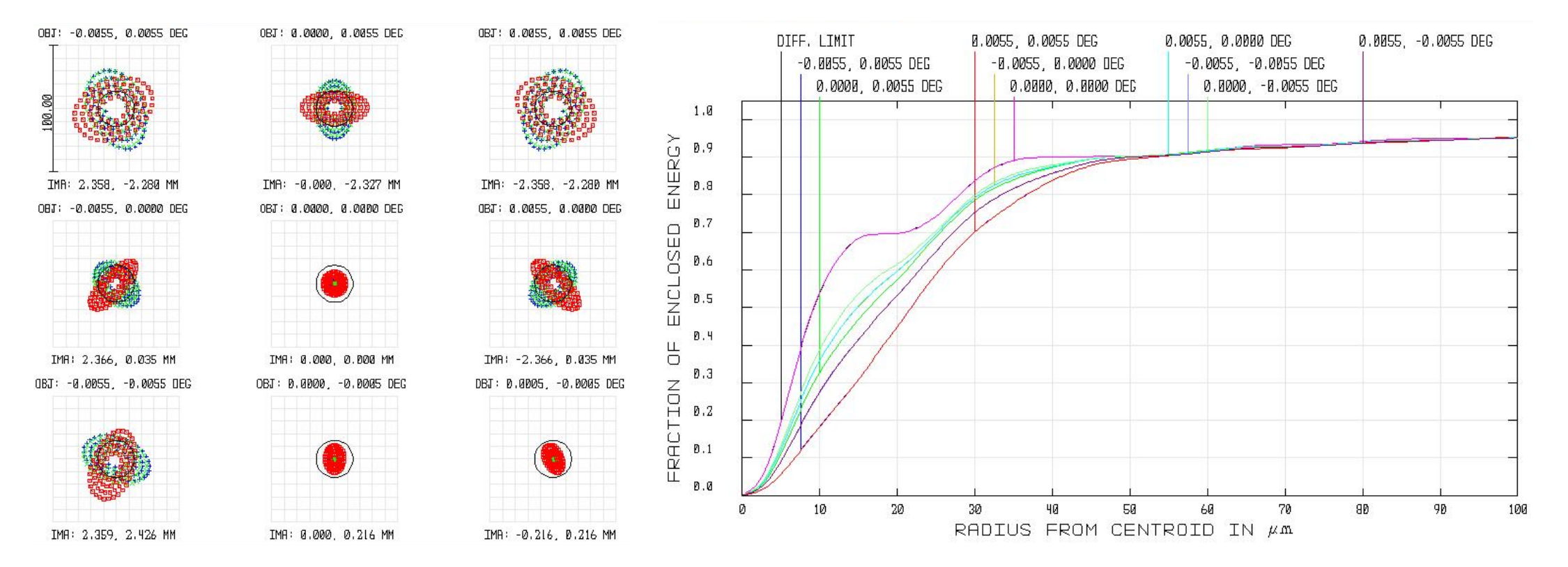}
   \end{tabular}
   \end{center}
   \caption[example] 
   { \label{fig:zemax_alpha} 
KAPAO-Alpha spot diagrams [left] and encircled energy curves [right] spanning a 40\arcsec x 40\arcsec field-of-view. As a prototype, KAPAO-Alpha was designed to have low-cost, off-the-shelf off-axis parabolas (OAPs) for the pupil relay optics. This allowed the undergraduate students to gain experience mounting and aligning OAPs, but came with limits to KAPAO-Alpha's optical performance. The spot diagrams [left] show that geometric aberrations due to our available optics become comparable to the diffraction limit, represented by the Airy disk (black circle).  FFT diffraction enclosed energy curves [right] show diffraction limited imaging along the optical axis and measurable deviations over the 40\arcsec x 40\arcsec field-of-view. This figure should be compared to Figure~\ref{fig:zemax_prime}, where we achieve a diffraction limited optical design for KAPAO-Prime over a much larger field of view.
 }
   \end{figure}

Figure~\ref{fig:alpha} presents schematic and as-built views of the KAPAO-Alpha optical table. Key components in the system were the aforementioned MEMS DM and WFS as well as an Andor iXon EMCCD optical imaging camera and off-the-shelf OAP relay mirrors. The iXon served as the visible light science camera, and to save time and expense pellicle beamsplitters were used for the WFS/science camera split and elsewhere. Overall throughput suffered, and was not optimized for this prototype system. Ultimately, the choice to design the relays to the Physik Instrumente piezo tip-tilt mirror (TT) and the MEMS DM by using only off-the-shelf OAPs led to a design that was within the diffraction limit for angles near the optical axis, (see Figure~\ref{fig:zemax_alpha}). Despite these known limitations, we were pleased with the ability to bring our students through Zemax and Solidworks design processes and get them familiar with mounting and aligning OAPs.

\graphicspath{{./images/}}

\graphicspath{{./images/}}

\begin{figure}[b]
   \begin{center}
   \begin{tabular}{c}
   \includegraphics[width=6.5in]{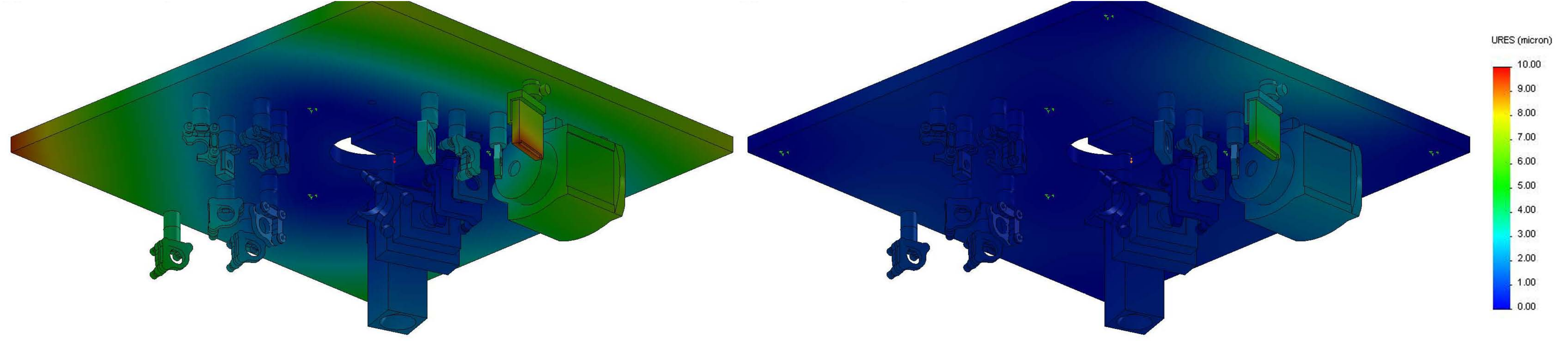}
   \end{tabular}
   \end{center}
   \caption[example] 
   { \label{fig:fea} 
Finite element analysis of the KAPAO-Alpha optical breadboard deflection with a 3-bolt [left] and 7-bolt [right] mounting pattern. The color scale shows deflections from 0 to 10 microns. With a flexure of no more than 6 microns and less than 1.5\% of an actuator at the MEMS DM, this student project gave us confidence in our mounting scheme. This is an example of the myriad different instrumentation design and construction tasks that our undergraduate students complete.
}
   \end{figure}

\begin{figure}
   \begin{center}
   \begin{tabular}{c}
   \includegraphics[width=6.5in]{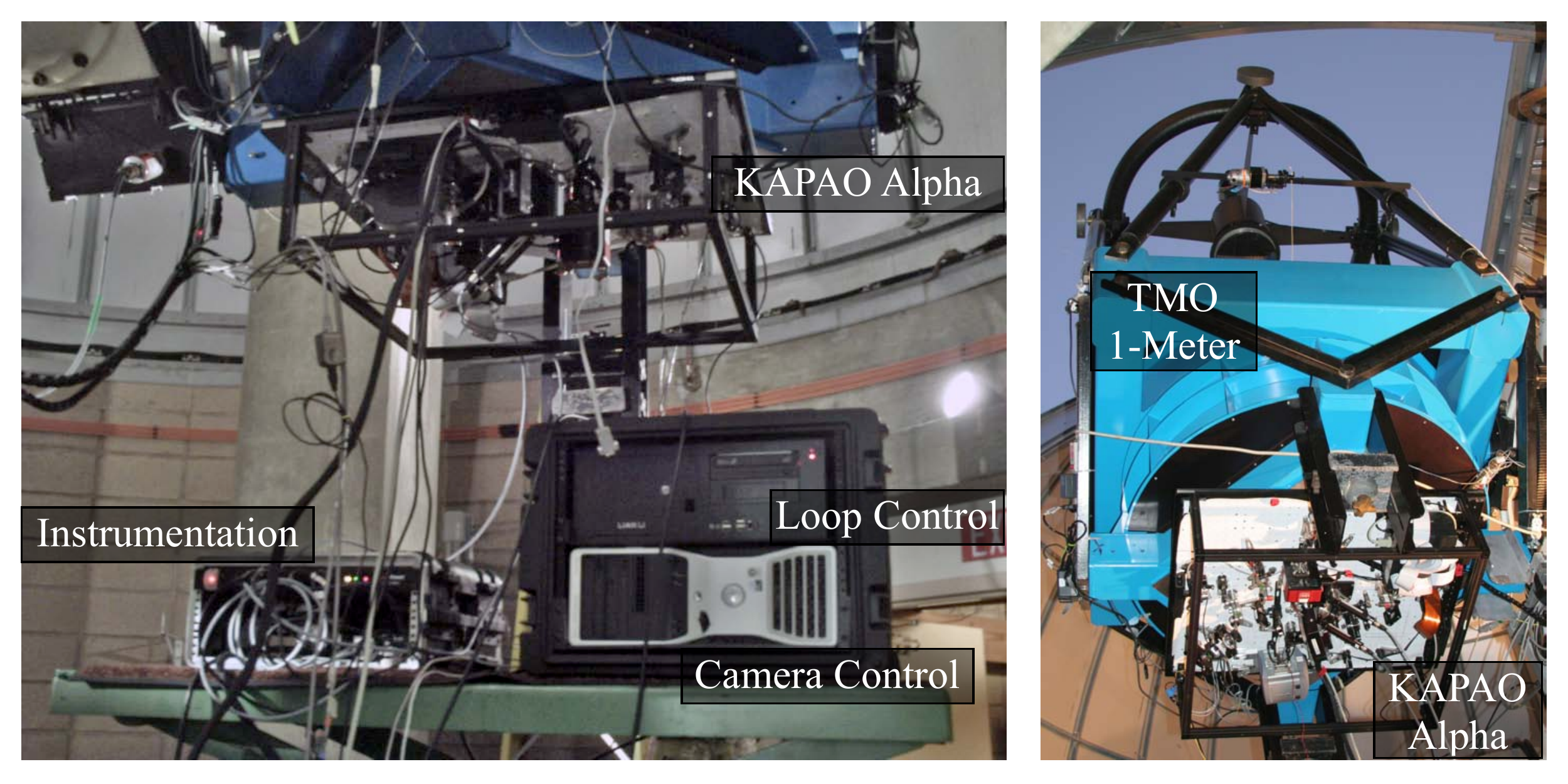}
   \end{tabular}
   \end{center}
   \caption[example] 
   { \label{fig:onscope} 
KAPAO-Alpha on the Table Mountain Observatory 1-meter telescope. The KAPAO-Alpha optical bench is shown from the side [left] mounted to the telescope and cabled to the MEMS DM and camera controllers [Instrumentation], which are further cabled to the Loop Control computer running the Robo-AO software under Linux and the Camera Control computer running our science and alignment cameras under Windows. Also shown [right], a view of the instrument mounted to the telescope pointed at zenith at dusk. The whole system is mounted onto the back of the telescope with the optics hanging down.
}
   \end{figure}

An example of one of the questions the prototype allowed us to answer was the mounting solution between KAPAO and the telescope. KAPAO-Alpha and -Prime are designed with a 1-meter square optical breadboard plate forming the basis of the instrument. Unlike Villages\cite{2008SPIE.7018E.126G}, KAPAO was designed without a spaceframe mount between the AO table and the telescope. Instead, in order to minimize the space taken up before the telescope back focal plane, we went with an ``upside-down" mounting scheme that placed the breadboard directly against the telescope primary mirror mount. Figure~\ref{fig:fea} shows a Finite Element Analysis of the resulting flexure and confirmed that such flexure could be minimized with appropriate mounting points. Figure~\ref{fig:onscope} shows the instrument mounted in just this manner, and we have had no issues with flexure.

\graphicspath{{./images/}}

\graphicspath{{./images/}}

\section{KAPAO-Alpha FIRST LIGHT} \label{sec:sections}
\label{sec:light}

With the rapid development made possible by our choice to construct the KAPAO-Alpha prototype system, we were able to get to the telescope in summer 2012. Figure~\ref{fig:onscope} shows image of the KAPAO-Alpha system mounted on the TMO 1-meter telescope and our required supporting electronics. Ultimately we have mounted a subset of the electronics adjacent to KAPAO-Alpha on the 1-meter, and placed the remaining electronics and control computers on racks in the dome connected via umbilicals. The mounting of the KAPAO-Alpha optics table to the telescope consists of a flip followed by a guided lift onto fiducial mounting pins. The system has made several trips between the Pomona College campus and the telescope and has proven to be robust and portable.

\graphicspath{{./images/}}
   \begin{figure}[h]
   \begin{center}
   \begin{tabular}{c}
   \includegraphics[width=3.75in]{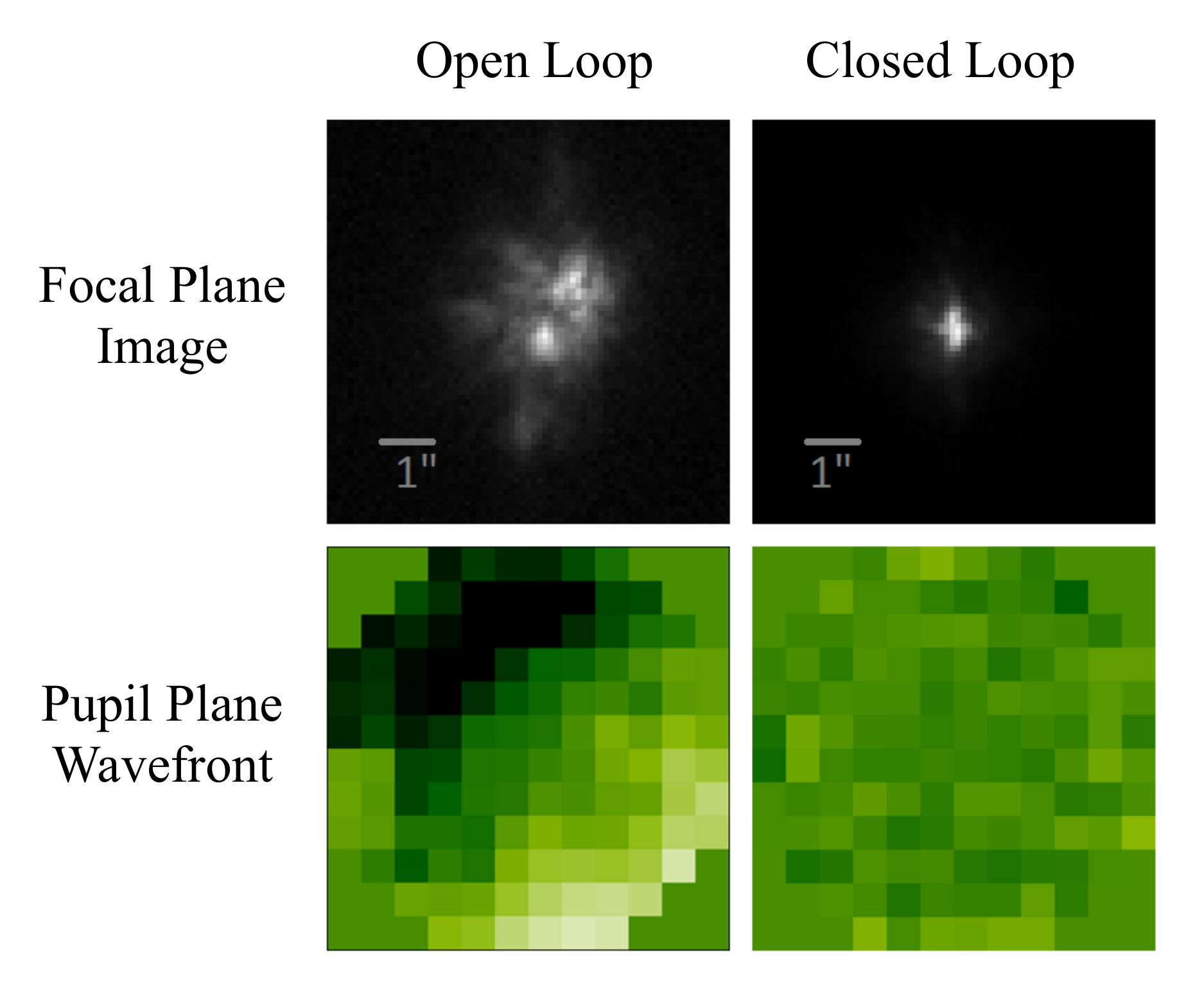}
   \end{tabular}
   \end{center}
   \caption[example] 
   { \label{fig:loop} 
KAPAO-Alpha instantaneous focal plane images [top] and phase maps [bottom] for open [left] and closed [right] loop operation on the bright star Deneb. A line denotes a scale of 1\arcsec. These visible-light images (without filter) show the dramatic reduction of wavefront error and the resulting image with an estimated Strehl of approximately 0.3.
}
   \end{figure} 

\graphicspath{{./images/}}
   \begin{figure}[h]
   \begin{center}
   \begin{tabular}{c}
   \includegraphics[width=3.75in]{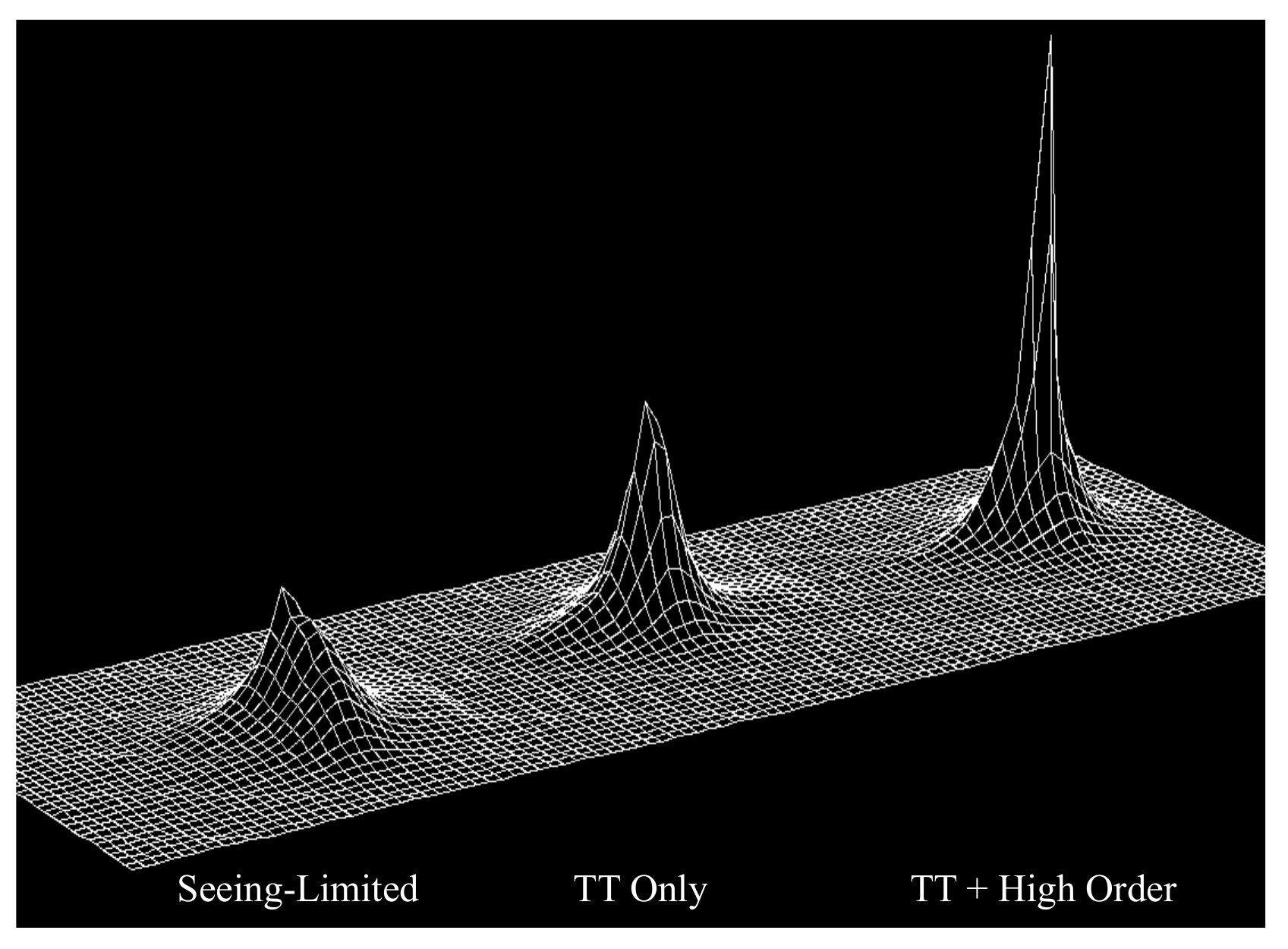}
   \end{tabular}
   \end{center}
   \caption[example] 
   { \label{fig:surface} 
KAPAO-Alpha on-sky closed-loop broadband optical observations of Deneb (v=1.5 mag). The surface plot shows the intensity distribution of an open-loop [left], tip-tilt (TT) only [middle], and closed-loop [right] images of Deneb. The seeing-limited image has a FWHM$>$1.5\arcsec, whereas corrected images have FWHM=1.0\arcsec and 0.5\arcsec, for TT-only and TT+high order correction, respectively.
}
   \end{figure} 

Figure~\ref{fig:loop} shows the results of closing the AO control-loop on the bright star, Deneb. The run, taken during atmospheric seeing of 1.5\arcsec, shows the capability of the system to correct static and time-varying aberrations. The closed loop image has a FWHM of 0.5\arcsec with an estimated Strehl of 0.3 in a broadband, unfiltered visible light image with the science camera. The Robo-AO control software is able to produce high-speed telemetry data of the loop performance and the lower half of Figure~\ref{fig:loop} shows a visualization of this data to show the pupil plane wavefront. Finally, Figure~\ref{fig:surface} presents a surface-plot view of the system in open-loop, tip-tilt only correction, and full AO correction modes.
The undergraduate student effort in all areas of this success of the KAPAO-Alpha instrument should be stressed. These include but are not limited to the aforementioned design and assembly, but also include creating data telemetry visualization and analysis pipelines.

\section{PLANS FOR KAPAO-Prime} 
\label{sec:prime}
\begin{figure}
   \begin{center}
   \begin{tabular}{c}
   \includegraphics[width=6.5in]{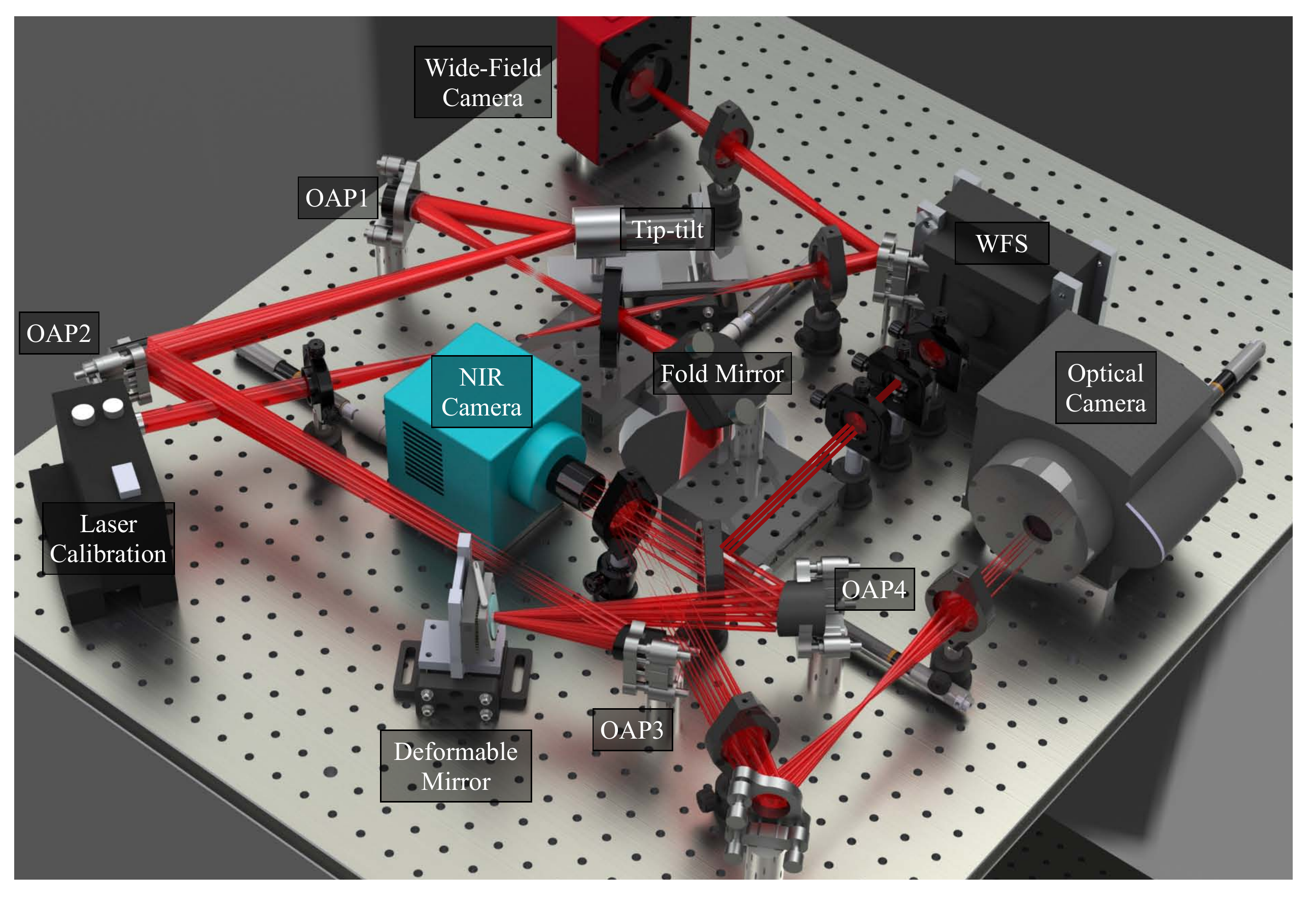}
   \end{tabular}
   \end{center}
   \caption[example] 
   { \label{fig:prime_layout} 	
A CAD Render of the KAPAO-Prime instrument design, showing the central Fold Mirror directing the telescope beam to the custom OAP mirrors, the Tip-Tilt mirror, the MEMS deformable mirror, the wavefront sensor, and science cameras. An Laser alignment source and Wide-Field camera are also evident. KAPAO-Prime is scheduled for first light later in 2013.
}
   \end{figure}

\begin{figure}
   \begin{center}
   \begin{tabular}{c}
   \includegraphics[width=6.5in]{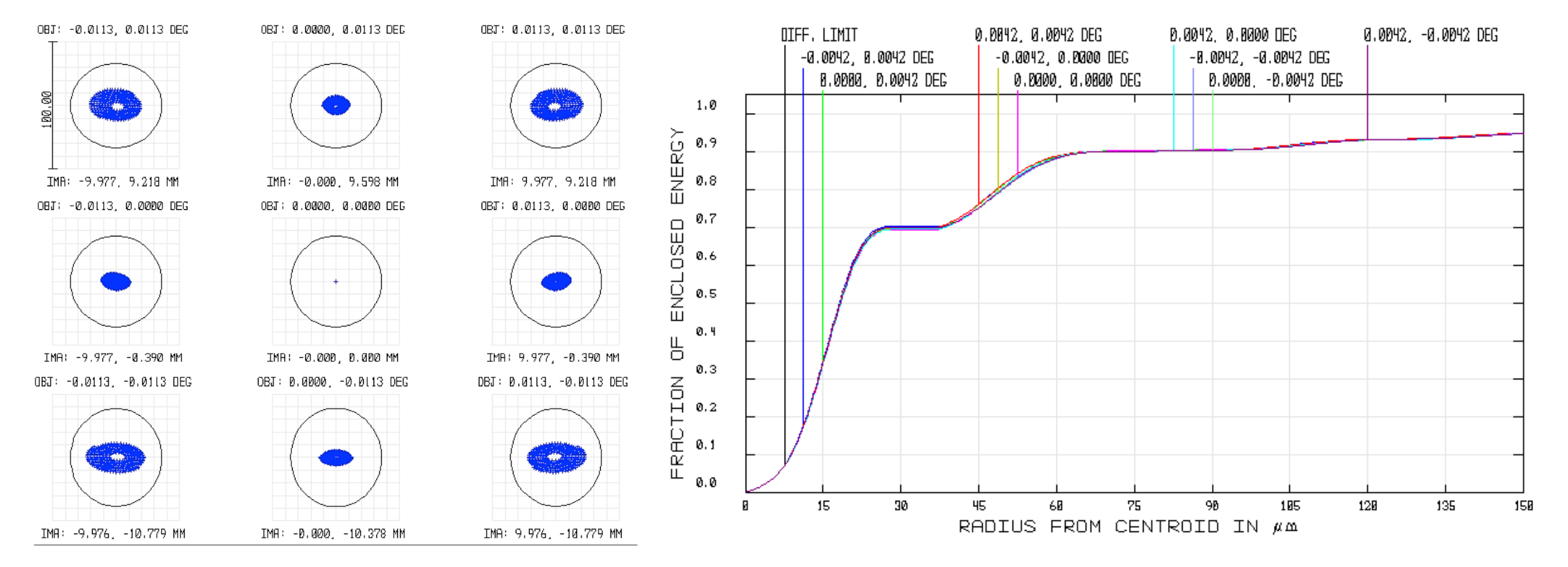}
   \end{tabular}
   \end{center}
   \caption[example] 
   { \label{fig:zemax_prime} 
KAPAO-Prime spot diagrams and encircled energy curves spanning a 80\arcsec x 80\arcsec field-of-view. The spot diagrams [left] show that geometric aberrations due to our optical design are negligible compared to the diffraction limit, represented by the Airy disk (black circle).  FFT diffraction enclosed energy curves [right] are indistinguishable from the diffraction limit over the central  80\arcsec x 80\arcsec FOV of the KAPAO-Prime instrument. This figure should be compared to Figure~\ref{fig:zemax_alpha}, where we used off-the-shelf OAPs in the design and construction of the KAPAO-Alpha prototype.
}
   \end{figure}

KAPAO-Prime is the facility class AO instrument for the TMO 1-meter telescope. KAPAO-Prime incorporates much of the basic system architecture from KAPAO-Alpha, a BMC MEMS 140-actuator DM, SciMeasure WFS, PI tip-tilt mirror, and an Andor EMCCD visible science camera. See Figure~\ref{fig:prime_layout}. KAPAO-Prime extends these features with the use of custom OAPs to provide far-superior optical design performance (see Figure~\ref{fig:zemax_prime}) as well as the use of dichroic beamsplitters and high-quality protected silver coatings to increase system throughput. Finally the system adds a near-infrared camera to provide dual-band imaging and high-strehl near-infrared imaging. First light for KAPAO-Prime is scheduled for later in 2013.

\graphicspath{{./images/}}

\graphicspath{{./images/}}

\section{Conclusions} 
\label{sec:conclusions}

At this stage in the KAPAO project we have designed, built and demonstrated a prototype AO system, using Robo-AO control software from our collaborators at CIT. We have a final design and are well underway in the construction of a capable facility instrument, KAPAO-Prime. KAPAO-Prime is a dual-band visible and near-infrared MEMS DM AO system. Undergraduate student involvement has been critical in every aspect of the project. The structure of our undergraduate programs at PC, SSU and HMC, encourages students to begin research early in their undergraduate career, and all of our programs have a senior thesis/capstone component that moves the students to authentic ownership of some portion of the project by the time of their graduation. KAPAO-Alpha is in current operation and KAPAO-Prime is scheduled for first-light later in 2013.
 
\appendix    

\acknowledgments     
 
This material is based upon work supported by the National Science Foundation under Grant No. 0960343. Additional funding for aspects of this work were provided by the Pomona College Summer Undergraduate Research Program, the Rose Hills Foundation (PC student funding), the Newkirk student assistantship (SSU student funding) and the Mt. Cuba Astronomical Foundation which supported the SSU AO testbed. Further acknowledgment should be made to early work in MEMS AO by the Villages team led by Don Gavel at the UCO/Lick Observatory Laboratory for Adaptive Optics and for the early work on the CIT Camera project by Matthew Britton. Co-authors Severson and Choi are also especially thankful for the excellent work of the student contributors to the KAPAO project and the key project assistance from Drs. Spjut, Baranec, and Riddle.




\bibliography{kapao}   
\bibliographystyle{spiebib}   

\end{document}